# Order In Chaos: Definite Rules That Govern The Drift Of Moon Away From The Earth

**Sohail Alam and Dr. B. K. Sharma**

***Department of Electronics and Communication Engineering***

***National Institute of Technology, Patna, India***

*Sa4250mnpo70@yahoo.com , electronics@nitp.ac.in*

## Abstract

When the Moon was formed it was much closer to the Earth than it is today. It just needed about 20 days then to go around the Earth. Now it takes the Moon 29.5 days to make one revolution. In order to follow the conservation of angular momentum the Moon had to either move closer to the Earth or recede from Earth. The data from the Lunar Laser Ranging Experiment confirms it to be moving away and the velocity of recession of the Moon have been found to be 3.8 cm/year. This rate is not constant though. At present the Moon's orbit has a radius of 384,000km.

But what is not yet known to all is that the orbital motion of the moon is actually a spiral motion. The moon is spiralling out and a very simple mathematical equation can describe the actual spiral motion of the Moon. The generalisation of this differential equation is the basic aim of this paper here.

## Introduction

The rock samples brought from our Moon during Apollo 11 to Apollo 17 Mission and during Luna 16 and Luna 20 Mission conclusively prove that Earth and Moon had been formed from the disc of accretion about 4.53 billion years ago but they were never a single body. The Age of Moon has been found incorrect by recent findings. The new age is around 4.467 billion years. Out of the many theories finally in 1984 at the International Conference held at Kona, Hawaii, Giant Impact Theory was accepted as the most consistent theory regarding Moon's origin.

When the Moon was formed it was much closer to the Earth than it is today. It just needed about 20 days then to go around the Earth. Now it takes the Moon 29.5 days to make one revolution. In order to follow the conservation of angular momentum the Moon had to either move closer to the Earth or recede from Earth. The data from the Lunar Laser Ranging Experiment confirms it to be moving away and the velocity of recession of the Moon have been found to be 3.8 cm/year. This rate is not constant though. At present the Moon's orbit has a radius of 384,000km.

The first laser ranging retro reflector was positioned on the Moon in 1969 by the Apollo 11 astronauts. By beaming laser pulses at the reflector from Earth, scientists have been able to determine the round-trip travel time that gives the distance between the two bodies at any time to accuracy of about 3 centimetres.

The reason for the increase is that the Moon raises tides on the Earth. Because the side of the Earth that faces the Moon is closer, it feels a stronger pull of gravity than the centre of the Earth. Similarly, the part of the Earth facing away from the Moon feels less gravity than the centre of the Earth. This effect stretches the Earth a bit, making it a little bit oblong. We call the parts that stick out "tidal bulges." The actual solid body of the Earth is distorted a few centimetres, but the most noticeable effect is the tides raised on the ocean. Now, all mass exerts a gravitational force, and the tidal bulges on the Earth exert a gravitational pull on the Moon. Because the Earth rotates faster (once every 24 hours) than the Moon orbits (once every 27.3 days) the bulge tries to "speed up" the Moon, and pull it ahead in its orbit. The Moon is also pulling back on the tidal bulge of the Earth, slowing the Earth's rotation. Tidal friction, caused by the movement of the tidal bulge around the Earth, takes energy out of the Earth and puts it into the Moon's orbit, making the Moon's orbit bigger (but, a bit paradoxically, the Moon actually moves slower!). The Earth's rotation is slowing down because of this. One hundred years from now, the day will be

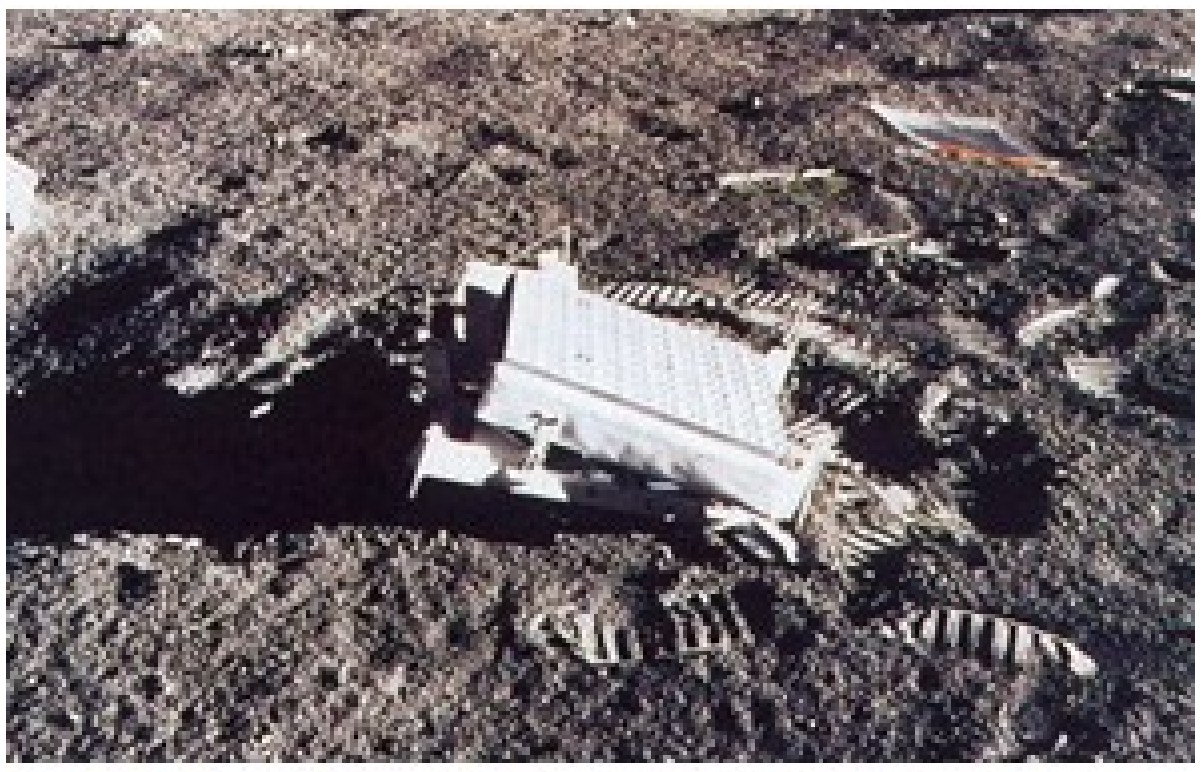

Fig1: Lunar Ranging Retro Reflector (LRRR)

2 milliseconds longer than it is now. This same process took place billions of years ago--but the Moon was slowed down by the tides raised on it by the Earth. That's why the Moon always keeps the same face pointed toward the Earth. Because the Earth is so much larger than the Moon, this process, called tidal locking, took place very quickly, in a few tens of millions of years. Many physicists considered the effects of tides on the Earth-Moon system. However, George Howard Darwin (Charles Darwin's son) was the first person to work out, in a mathematical way, how the Moon's orbit would evolve due to tidal friction, in the late 19th century. He is usually credited with the invention of the modern theory of tidal evolution.

There is a time lag between Earth-Moon interaction and Earth's deformation. It is this time lag and belated deformation of Earth which causes a retarding torque on Earth. The two bulges act as brake-shoes as these brake-shoes act in Cycle wheel.

# Problem Definition

It was established a century ago by George Howard Darwin that Earth-Moon interaction is causing Lunar Orbital Radius expansion and Secular Lengthening of Day (L.O.D). This paper intends to study the generalisation of the theoretical formulation of Lunar Orbital Radius Expansion.

# Details about the Problem

When a capacitance is being charged from a constant voltage source using a resistance, the capacitance voltage exponentially grows or decays with the time constant of growth or decay given by T = RC. Is there some such formulation in the orbital radius expansion or contraction curve? Is there a similar time constant here? And if indeed there is time constant what does it depend upon? By the studies conducted till now we have defined a time constant $T = \frac{a_{G2}-a_{G1}}{V_{max}}$ where $a_{G1}$ is inner geosynchronous orbit, $a_{G2}$ is outer geosynchronous orbit and $V_{max}$ is the maximum velocity of recession as a result of gravitational sling shot effect.

But what is not yet known to all is that the orbital motion of the moon is actually a spiral motion. The moon is spiralling out and a very simple mathematical equation can describe the actual spiral motion of the Moon. The generalisation of this differential equation is the basic aim of this paper here. It has been shown in the thesis of Dr. Bijay Kumar Sharma, named as "THE DYNAMICS OF PLANETARY SATELLITES AND THE GENERALIZATION OF THE SAME PROVIDES A NEW THEORY OF THE BIRTH AND EVOLUTION OF OUR SOLAR SYSTEM" that in any two body system there are two Geo-synchronous orbits $a_{G1}$ and $a_{G2}$.

When the secondary is tidally locked with the primary as all the natural satellites are, then they are called synchronous motion. Moon is in synchronous motion. Hence we see the same face of Moon eternally. Whereas, Charo is tidally interlocked with Pluto. This is analogous to geo-synchrony. Charon spin period= Charon's orbital period= Pluto's spin period. If Moon was tidally interlocked with Earth then Moon would be permanently fixed over a given point say Patna (the capital of Bihar, INDIA) as a geo-synchronous satellite is. Plus when secondary body is in sub-synchronous orbit, angular momentum is transferred from Moon to Earth (secondary to primary body). Therefore Moon gets trapped in a contracting spiral orbit. If it was in extra synchronous orbit then angular momentum would be transferred from Earth to Moon as presently it is happening and Moon is receding at the rate of 3.8cm.The Geo-Synchronous orbits may be defined as that where the spin period of the secondary body equals that of the Primary as well and the orbital period of the Secondary equals the spin period of the Primary. If the accreting body or the captured body, as is the case for Martian Satellite Phobos, lies within $a_{G1}$ then the Satellite is gravitationally launched on an inward spiral path towards its sure doom. If the accreting zone lies beyond $a_{G1}$ as is the case with our Moon or the captured body lies beyond $a_{G1}$ as is the case for the Martian Satellite Deimos, then the said satellite experiences an impulsive, gravitational sling shot effect because of Conservation of Energy Principle.

It has rightly been assumed by Dr. B.K. Sharma that the trajectory of evolution obeys the rules analogous to Fourier expansion series having the form $\mathbf{A}e^{-\frac{t}{3T}} + \mathbf{B}e^{-\frac{t}{2T}} + \mathbf{C}e^{-\frac{t}{T}} + \cdots$ where **A, B, C** … are some constants. Here we approximate the differential equation to be of third order only, (it can be of higher order but the mathematics get quite involved above order eight or nine, but the basic formulation is the same).

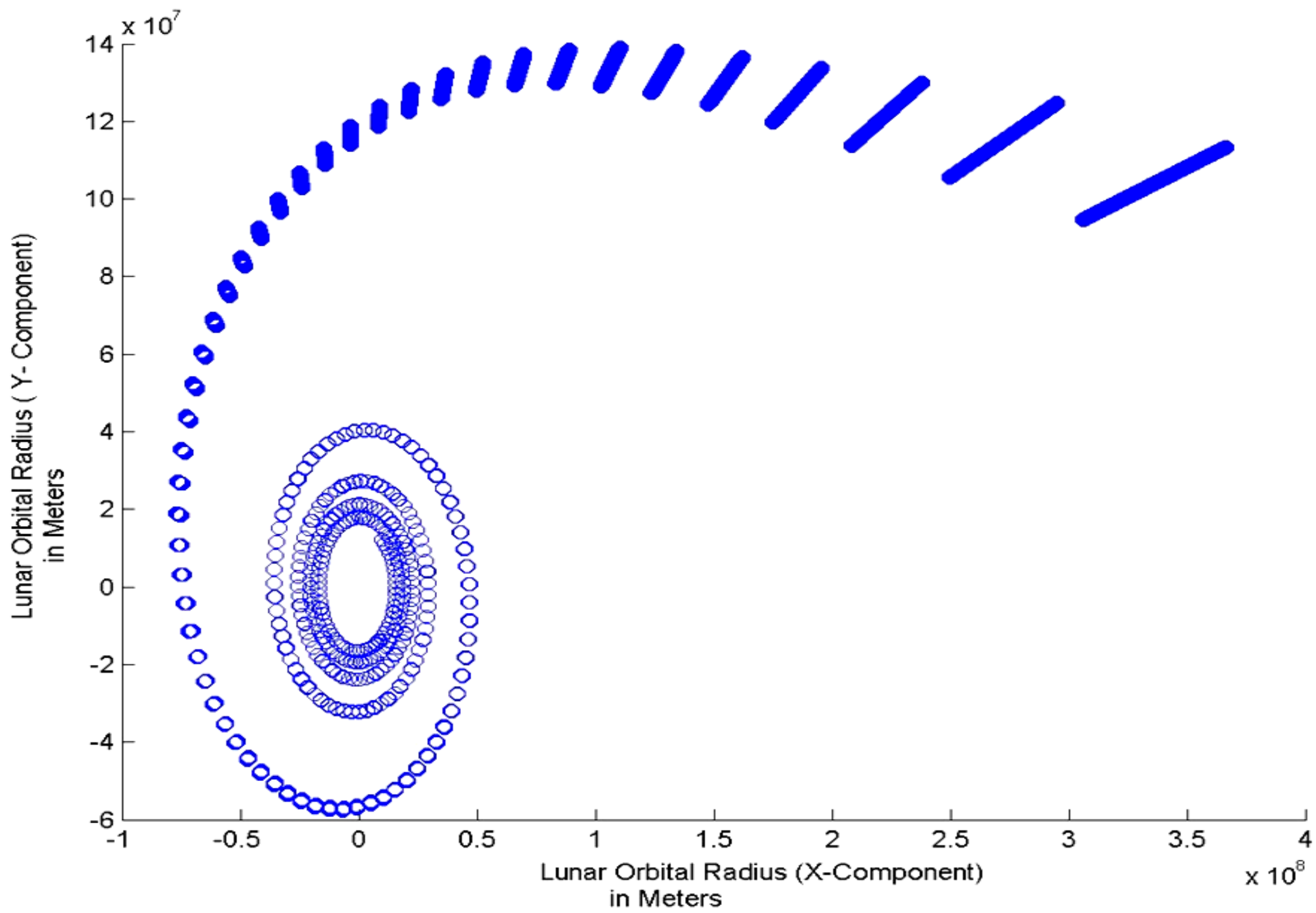

The Figure represents the Spiral Trajectory of Moon traversed from the birth to the present Orbit of 3.844x10^8m over 4.53Gy of time period. (Ref. 4)

***All these mathematical formulation has been done using Wolfram Mathematica 7 software.***

Let us take the case of a polynomial whose roots are -1/T, -1/2T and -1/3T.

$$\textbf{Polynomial} = \left(x+\frac{1}{T}\right)\left(x+\frac{1}{2T}\right)\left(x+\frac{1}{3T}\right)$$

On expansion using the **Expand** command we get the following result

$$\frac{1}{6T^3}+\frac{x}{T^2}+\frac{11x^2}{6T}+x^3$$

Now observing the above expression we construct a differential equation as follows

$$\frac{d^3x(t)}{dt^3}+\frac{11}{6T}\frac{d^2x(t)}{dt^2}+\frac{1}{T^2}\frac{dx(t)}{dt}+\frac{1}{6T^3}x(t)=0$$

Consider $\mathbf{r_L}$ to be the radial distance of the Moon from Earth and $\mathbf{a_{g1}}$ and $\mathbf{a_{g2}}$ are the first and second Geo-synchronous orbit respectively. Then consider

$$\mathbf{x}=\frac{\mathbf{a_{g2}}-\mathbf{r_L}}{-\mathbf{a_{g1}}+\mathbf{a_{g2}}}$$

Now solving the differential equation using **DSolve** command

$$\mathbf{DSolve}[\mathbf{x}'''[\mathbf{t}] = \\ = -\mathbf{11}/(\mathbf{6T})\,\mathbf{x}''[\mathbf{t}] - \mathbf{1}/\mathbf{T}^2\,\mathbf{x}'[\mathbf{t}] \\ - \mathbf{1}/(\mathbf{6T}^3)\,\mathbf{x}[\mathbf{t}], \mathbf{x}[\mathbf{t}], \mathbf{t}]$$

The solution is

$$\frac{\mathbf{a_{g2}}-\mathbf{r_L}}{-\mathbf{a_{g1}}+\mathbf{a_{g2}}}=\mathbf{C}[\mathbf{1}]e^{-\frac{t}{3T}}+\mathbf{C}[\mathbf{2}]e^{-\frac{t}{2T}}+\mathbf{C}[\mathbf{3}]e^{-\frac{t}{T}}$$

where C[1], C[2], C[3] are the three constants of integration.

This solution is exactly what we have assumed to be.

There are four unknowns here C[1], C[2], C[3], T

Also any of the following boundary conditions may be used

$r_L = a_{G2}$ at t=infinity and

$r_L = a_{G1}$ at t= 0

And in case of Earth-Moon system the two Geo-synchronous orbits have been found to be

$a_{g1}= 1.46257569*10^7$ m

$a_{g2}= 5.52887891*10^8$ m

We know from Laser Lunar Ranging Experiment that the Moon is at a distance of $3.844*10^8$ m from the Earth at present, after $4.53*10^9$ years after the Giant Impact.

Hence, $r_L= 3.844*10^8$ **at** $t= 4.53*10^9$

---

# Results and discussion

All we have to do is find the constants of integration and the Time Constant of Evolution. The Time Constant was found out to be $T=3.90028\times10^9$, by solving the following equation

$$\mathbf{Solve}\left[\mathbf{1}-e^{-4.53\times10^9/T}==\mathbf{0.68697},\mathbf{T}\right]$$

Now satisfying different points in space at which the Moon was present at time t (after the Giant Impact), we get simultaneous equations containing the constants of integration C[1], C[2], C[3].

For example satisfying the following points $r_L = 2.227*10^8$ at $t=1.73*10^9$, $r_L= 3.844*10^8$ at $t= 4.53*10^9$, and $r_L = a_{G1}$ at t= 0 and solving for C[1], C[2], C[3] we get

C[1]= 1.88839, C[2]= -2.72107, C[3]= 1.83268

Now putting these values of T, C[1], C[2] and C[3] and plotting the radial distance with respect to time we get the following curve:

Also on plotting the different points in space at any given time where the moon would have been possibly present using the data generated by Prof. B.K.Sharma we get the experimental curve as follows:

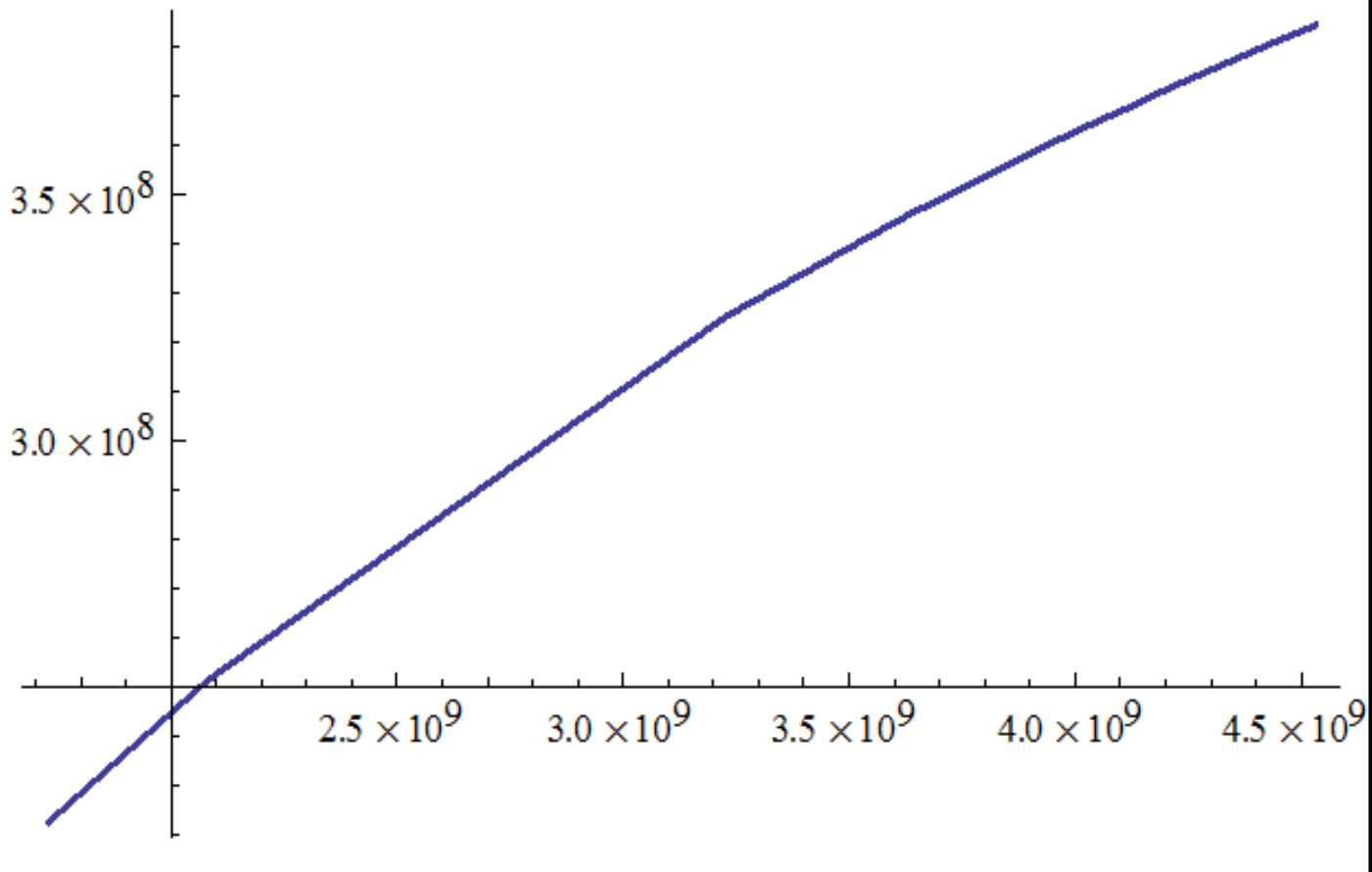

Fig.3: Radial Distance V/s Time (Using the generated data)

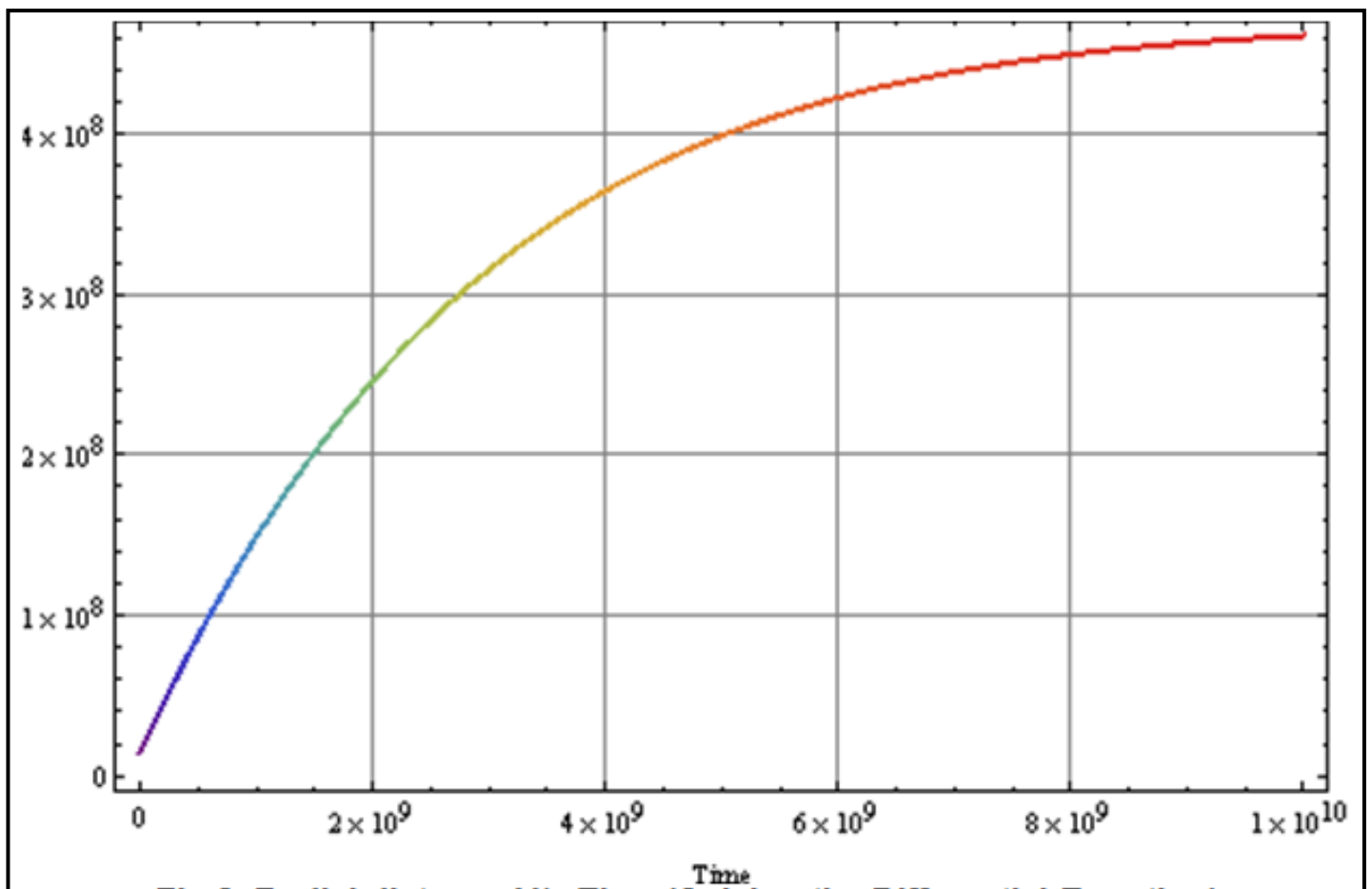

Fig.2: Radial distance V/s Time (Solving the Differential Equation)

## Summary

It can be seen that the moon actually is moving outward and is approaching for its outer geo-synchronous orbit. Also when the two plotted graphs are compared we find a very accurate match.

We need to study the orbital radius expansion or contraction and see if they fit in a general formulation. Also these formulations will the give some very significant idea about the Lengthening of Day (L.O.D).

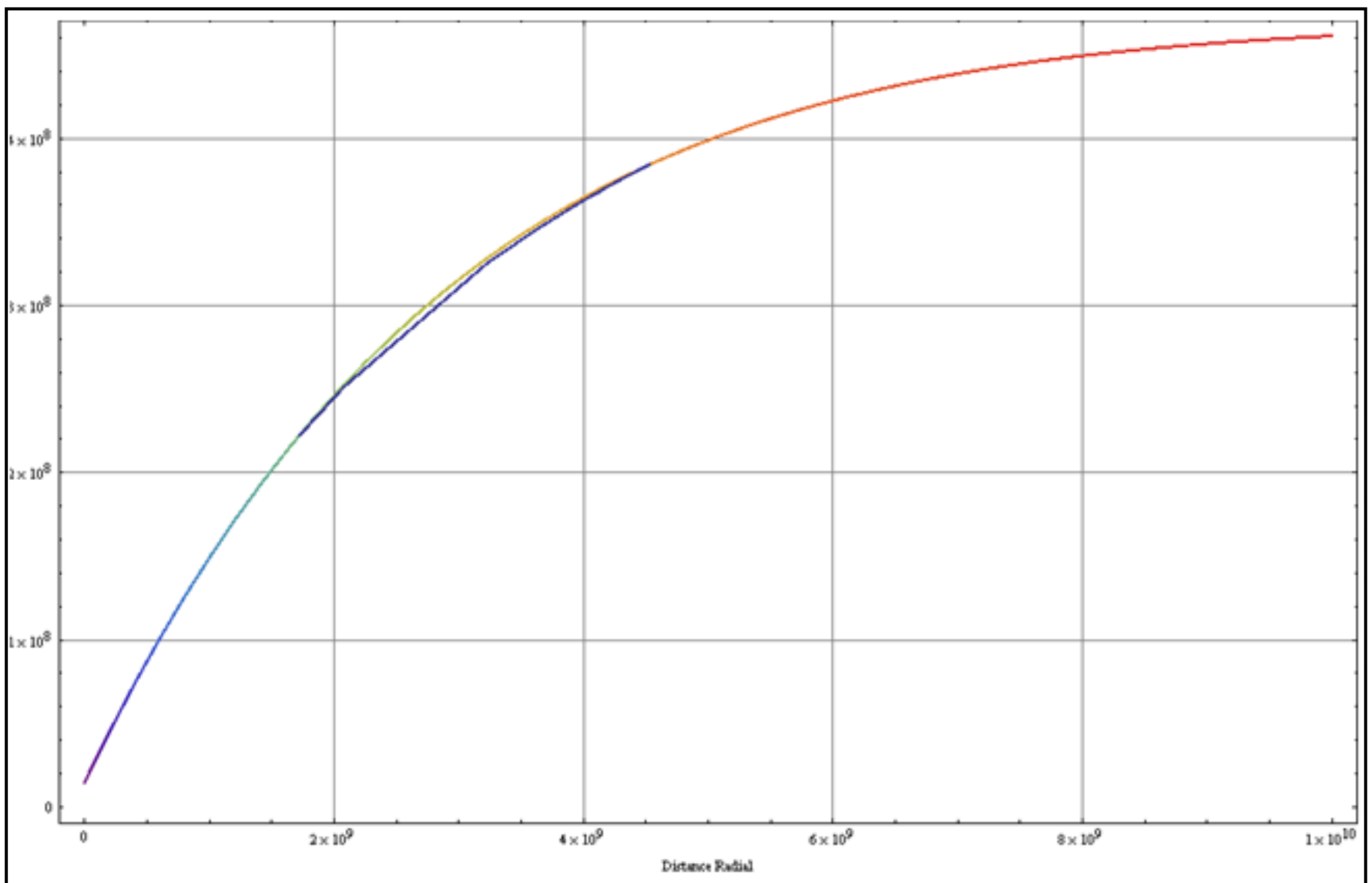

Fig. 4 Comparing the two plots

## Future Directions

There are several causative factors for the deviations in the actual sidereal day L.O.D. Curve with respect to the theoretical L.O.D. curve as predicted on the basis of Earth-Moon Tidal Interaction. Because it has a causative factor hence the deviations are not random but chaotic. If they are indeed chaotic they should be carrying the signatures of the various causative factors such as plate-tectonics, ice-caps expansion and thawing, electromagnetic coupling between core and mantle, El-Nino Ocean Currents and global wind pattern interactions with various mountain ranges. If the various signatures are identified then we may isolate the precursors of Earth-quake and sudden Volcanic Eruptions in L.O.D. curve fluctuations. With the advent of quantum gyro developed at University of California, Berkeley, it should not be difficult to identify the fluctuations which characterize the plate-tectonic movements. These fluctuations will be categorized according to the magnitude of tremors. Certain categories will act as the precursors of Earthquakes also.

We must also study the orbital radius expansion or contraction of satellites and see if they fit in a general formulation. Among the natural satellites we will cover Moon, Phobos, Deimos, Charon, Titan and a few other large sized satellites.

---------------

……..